\documentclass{article}

\usepackage{arxiv}

\usepackage[utf8]{inputenc} % allow utf-8 input
\usepackage[T1]{fontenc}    % use 8-bit T1 fonts
\usepackage{hyperref}       % hyperlinks
\usepackage{url}            % simple URL typesetting
\usepackage{booktabs}       % professional-quality tables
\usepackage{amsfonts}       % blackboard math symbols
\usepackage{nicefrac}       % compact symbols for 1/2, etc.
\usepackage{microtype}      % microtypography
\usepackage{lipsum}		% Can be removed after putting your text content
\usepackage{graphicx}
\usepackage{natbib}
\usepackage{doi}
\usepackage{tikz}
\usepackage{listings}
\usepackage{float}
\usepackage{comment}

\usetikzlibrary{shapes.geometric, arrows}
\usetikzlibrary{positioning}

\tikzstyle{startstop} = [rectangle, rounded corners, minimum width=3cm, minimum height =1cm,text centered,
text width=3cm,  draw=black, fill=red!30]
\tikzstyle{io} = [trapezium, trapezium left angle=70, trapezium right angle=110, minimum width=1cm,
text width=3cm,  minimum height=1cm, text centered, draw=black, fill=blue!30]
\tikzstyle{process} = [rectangle, minimum width=3cm,
text width=3cm,  minimum height=1cm, text centered, draw=black, fill=orange!30]
\tikzstyle{decision} = [diamond, minimum width=3cm,
text width=3cm,  minimum height=1cm, text centered, draw=black, fill=green!30]
\tikzstyle{arrow} = [thick,->,>=stealth]
\tikzstyle{request} = [rectangle, rounded corners, minimum width=3cm, minimum height=1cm,text centered,
text width=3cm,  draw=black, fill=green!30]

\title{Automatically detecting scientific political science texts from a large general document index}

\author{ Nina Smirnova\\
    GESIS -- Leibniz Institute for the Social Sciences\\
    Unter Sachsenhausen 6-8 \\
    50667 Köln \\
    \texttt{nina.smirnova@gesis.org}
}

\renewcommand{\headeright}{Technical Report}
\renewcommand{\undertitle}{Technical Report}
\renewcommand{\shorttitle}{}

\hypersetup{
pdftitle={Detecting texts from the political science domain},
pdfsubject={cs.DL},
pdfauthor={Nina Smirnova},
pdfkeywords={Classification,  Political Science, Pollux, GESIS, Python},
}

\begin{document}

\maketitle

\begin{abstract}

This technical report outlines the filtering approach applied to the collection of the Bielefeld Academic Search Engine (BASE) data to extract articles from the political science domain. We combined hard and soft filters to address entries with different available metadata, e.g. title, abstract or keywords. The hard filter is a weighted keyword-based approach. The soft filter uses a multilingual BERT-based classification model, trained to detect scientific articles from the political science domain. We evaluated both approaches using an annotated dataset, consisting of scientific articles from different scientific domains. The weighted keyword-based approach achieved the highest total accuracy of 0.88. The multilingual BERT-based classification model was fine-tuned using a dataset of 14,178 abstracts from scientific articles and reached the highest total accuracy of 0.98. The proposed filtering approach can be applied for filtering metadata from other scientific domains and therefore improve the overview of the domain-related literature and facilitate efficiency in research. 
\end{abstract}

% keywords can be removed

%\section*{Disclaimer}
%\label{sec:disclaimer}

\section{Introduction}

The specialized information service (FID) Political Science - Pollux \footnote{\url{https://www.pollux-fid.de/}} aims to offer political scientists subject-specific and fast access to special literature and information relevant to their research. Pollux obtains data from different providers, one of them is BASE \footnote{\url{https://www.base-search.net/}} (Bielefeld Academic Search Engine), which indexes the metadata of all kinds of academically relevant resources. Many BASE entries have no domain classification,  therefore we developed a filtering approach which can detect entries pertinent to political science from the huge collection of BASE metadata. Additionally, publications relevant to political science also might be included in other scientific domains. State and University Library Bremen (SuUB) has used BASE filtering methods for their Electronic library (E-LIB) since 2012 \citep{Blenkle:810613, BlenkleNoelte2017}. The core BASE filtering approach including a weighted keyword-based approach was developed in 2016 by the SuUB team for the Pollux project. 
%This approach is a basis for filtering methods for FID Pharmacy~\footnote{\url{https://www.tu-braunschweig.de/en/ub/specialised-information-service-pharmacy}} and FID Education~\footnote{\url{https://www.fachportal-paedagogik.de/en/literatur/produkte/fachinformationsdienst/projektinformation.html}}. 

In this technical report, we describe the training procedure, training data and usage of two classification models which are part of one of the modules (soft filter) of our filtering approach (see Figure~\ref{fig:workflow_base}). Additionally, we provide an evaluation of both classification models and a weighted keyword-based approach (hard filter) on a specially designed test dataset.

\subsection{BASE}
\begin{figure}[h!]
\centering
  \includegraphics[width=1\textwidth]{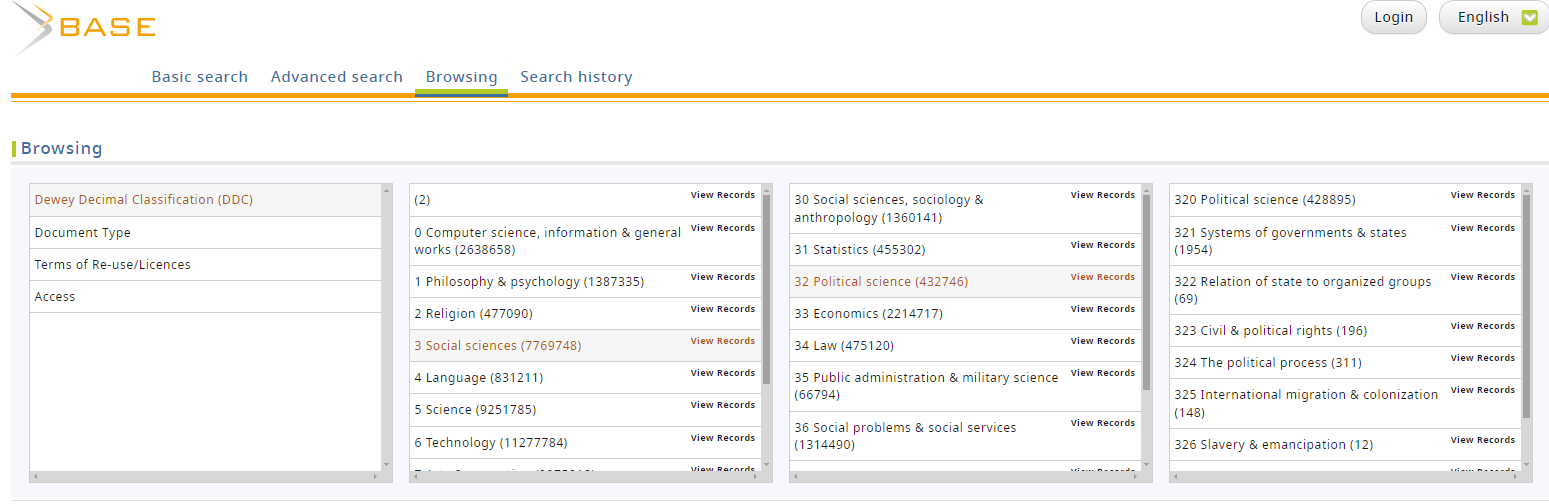}
  \caption{Example of the BASE interface and metadata}
  \label{fig:base_metadata}
  %\vspace{-4mm}
\end{figure}

The BASE is a search engine for academic web resources operated by Bielefeld University Library. The BASE provides over 340 million documents from different content providers and consists of various kinds of academic resources, e.g., journals, books digital collections, etc \footnote{\url{https://www.base-search.net/about/en/index.php?}}. Many BASE entries are classified by the Dewey Decimal Classification (DDC). The DDC is a system for organizing library contents by dividing all knowledge into ten categories, each assigned a range of 100 numbers. Aside from DDC BASE data has many other important metadata, such as source, journal name, authors' names, journal name, etc. Figure~\ref{fig:base_metadata} shows the BASE base interface and basic metadata.

%71,209,199 - relevant data
%100,404,711 - total count

\subsection{Filtering workflow}
Developing approaches for filtering political science articles from a collection of unlabeled or multidisciplinary scientific articles is crucial for several reasons. \cite{holyst_protect_2024} argue that excessive information constrains our ability to assess information and make optimal decisions. 
Therefore, excluding articles irrelevant to political science can enhance political scientists' time efficiency and enable focusing on the materials most pertinent to their work. 
Political science often intersects with other disciplines. Finding articles related to politics in other disciplines will provide political researchers with opportunities for interdisciplinary collaboration, enrich their perspectives and enhance the depth of their research. Political science is a dynamic field, with new events, policies, and trends continuously emerging. Timely access to relevant literature allows political researchers to remain abreast of the latest developments in the political field, enabling informed analyses and discussions. 

\begin{figure}[h!]
\centering
  \includegraphics[width=0.8\textwidth]{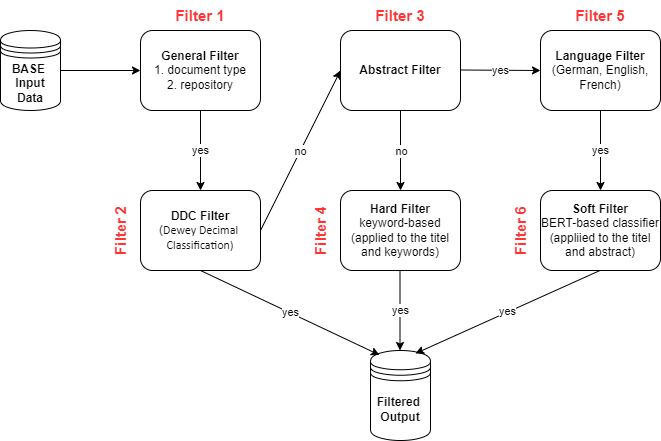}
  \caption{Pipeline for the BASE filtering approach}
  \label{fig:workflow_base}
  %\vspace{-4mm}
\end{figure}

The filtering approach for the BASE data collection comprises different modules to address entries with different metadata available as illustrated in  Figure~\ref{fig:workflow_base}.
Initially, the data is filtered according to parameters such as data type and source. Subsequently, in the second filter articles already categorized under the DDC system as political science (DDC 320-328) are selected. The third filter selects articles with a valid abstract. For articles lacking abstracts, a keyword-based filter (hard filter) is applied. Articles containing abstracts undergo a language filter, and those written in permissible languages are then subjected to a BERT-based classification model (soft filter). 

\begin{figure}[h!]
\centering
  \includegraphics[width=0.8\textwidth]{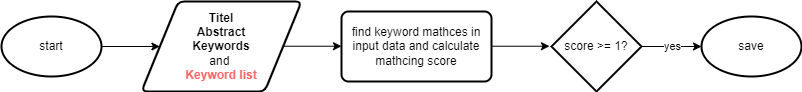}
  \caption{Soft filter simplified workflow}
  \label{fig:workflow_keywords}
  %\vspace{-4mm}
\end{figure}

The two major modules of this approach are so-called hard and soft filters. The hard filter is a weighted keyword-based filter approach. We developed a set of keywords which we believe to be effective for identifying data related to the political science domain. Each keyword was manually assigned with a score. The score is assigned according to the keyword's relevancy for the domain, as Table~\ref{tab:keywords_example} demonstrates. Figure~\ref{fig:workflow_keywords} shows a simplified workflow for the soft filter. The filtering algorithm looks for the keywords in the metadata and returns a sum of scores of all found keywords. If this score is more or equal to 1 the article is marked as relevant to the domain. This module is applied to the entities which do not have a full abstract, and therefore the filter is applied to the entity’s title and keywords if the latter are available. 

\begin{table}[h!]
\centering
    \begin{tabular}{cc}
    \toprule
         keyword& score\\
    \midrule
         *politik& 1\\
         bürgerkrieg& 0.6\\
         policy& 0.4\\
    \bottomrule
    \end{tabular}
    \caption{Example of a keyword list for the hard filter}
    \label{tab:keywords_example}
\end{table}

The soft filter uses a BERT-based classification model, trained to detect scientific articles from the political science domain. The model was finetuned on abstracts from scientific articles, therefore the filter is applied to the entity’s title and abstract.

\section{Soft Filter}

For our soft-filter approach, we trained two classification models: English and multilingual, which can detect texts from the political science domain. 

\subsection{English classification model}

The English classification model to detect texts from the political science domain (SSciBERT\_politics)\footnote{\url{https://huggingface.co/kalawinka/SSciBERT_politics}} is based on the SSCI-SciBERT \citep{shen_sscibert_2023} model. SSCI-SciBERT was trained on the abstracts of social science research articles, therefore, we opted to use this model.
The model was fine-tuned using a dataset of 2,919 abstracts from scientific articles retrieved from the BASE and POLLUX collections. 

The SSciBERT\_politics was initially integrated into the soft filter of the BASE filtering pipeline in the POLLUX infrastructure. Nevertheless, the model can be used separately with the Transformers library~\footnote{\url{https://huggingface.co/docs/transformers/en/index}}, as Listing~\ref{listing:SSciBERT-politics} demonstrates.  

\begin{minipage}{\linewidth}
\begin{lstlisting}[caption = Usage example of SSciBERT\_politics with the Transformers library, label = listing:SSciBERT-politics]
from transformers import AutoTokenizer, AutoModelForSequenceClassification
from transformers import pipeline

tokenizer = AutoTokenizer.from_pretrained('kalawinka/SSciBERT_politics')
model = AutoModelForSequenceClassification.from_pretrained('kalawinka/SSciBERT_politics')
pipe = pipeline("text-classification", model=model, tokenizer = tokenizer, max_length=512, truncation=True)

pipe('add scientific abstract')
\end{lstlisting}
\end{minipage}

Applying this model to the abstract of the technical report yields the output presented in Listing~\ref{SSciBERT-politics-out}.

\begin{minipage}{\linewidth}
\begin{lstlisting}[caption = Example of SSciBERT\_politics output, label = SSciBERT-politics-out]
    [{'label': 'multi', 'score': 0.9991981387138367}]
\end{lstlisting}
\end{minipage}

The model was fine-tuned utilizing the Transformers library and using the AdamW optimizer\footnote{\url{https://huggingface.co/docs/bitsandbytes/main/en/reference/optim/adamw}}. Evaluation was done at the end of each epoch. The total number of training epochs to perform was set to 4. The initial learning rate was set to 5e-5. The training was performed using one NVIDIA A40 GPU.

Listing~\ref{listing:SSciBERT-politics-training} demonstrates the training parameters used to fine-tune the SSciBERT\_politics model.

\begin{minipage}{\linewidth}
\begin{lstlisting}[caption = Training parameters for the finetuning  of SSciBERT\_politics with the Transformers library, label = listing:SSciBERT-politics-training]
training_args = TrainingArguments(output_dir=output_dir, 
                                  evaluation_strategy="epoch",
                                  num_train_epochs = 4,
                                  )
\end{lstlisting}
\end{minipage}

\subsubsection{Training data}\label{training_data_english}

\begin{figure}[H]
\centering
  \includegraphics[width=1\textwidth]{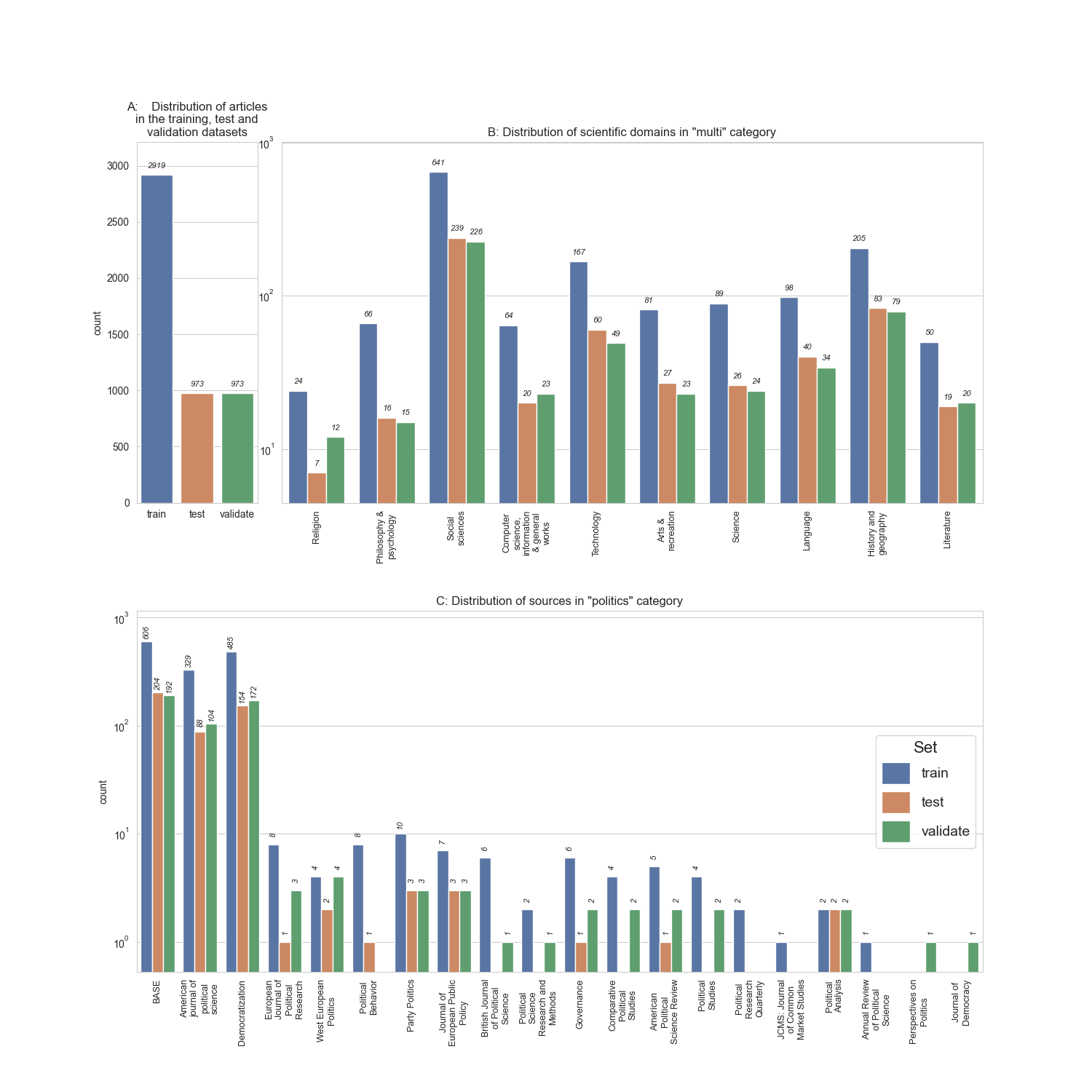}
  \caption{Distribution of training data in English corpus}
  \label{fig:corpus1_combined}
  %\vspace{-4mm}
\end{figure}

Due to a lack of manually annotated data, we employed a semi-automated method to generate the labelled training dataset. Figure~\ref{fig:corpus1_combined}-A illustrates the distribution of articles across the training, test, and validation corpora. Accordingly, the test and validation datasets each contain 20\% of the data, while the training corpus comprises 60\% of the total data. 
Data from the BASE and POLLUX collections were utilized to construct the training corpus.  
To acquire scientific abstracts within the domain of political science, we utilized both POLLUX and BASE collections. We compiled a list of political science journals in English. Afterwards, we collected articles from the POLLUX collection published in these journals containing an abstract and published after 2018. Following, we collected articles from the BASE collection adhering to specific criteria. Only articles with the types article or review, DDC classification 320-328 (Political science), written in English, containing an abstract and published after 2018 was selected. Figure~\ref{fig:corpus1_combined}-C shows the distribution of sources in the 'politics' category. To collect data from other (not political science) domains, we utilized only articles from the BASE collection. We sampled articles which correspond to the following criteria. Only articles with the types article or review, DDC classification not 320-328 (Political science), written in English containing an abstract and published after 2015 were selected. Additionally, we designed our query to restrict the selection only to articles which do not contain the keyword 'politi*' in title, keywords or abstracts. Figure~\ref{fig:corpus1_combined}-B shows the distribution of scientific domains in the 'multi' category\footnote{One article can belong to multiple scientific domains.}. After collecting the data from both collections, we verified the language of the abstracts to ensure they were in English and excluded abstracts containing fewer than 20 words.  

%\subsubsection{Training}

\subsection{The multilingual classification model}

The multilingual classification model to detect texts from the political science domain (bert-base-ml-politics)\footnote{\url{https://huggingface.co/kalawinka/bert-base-ml-politics}} is based on BERT multilingual base model \citep{devlin_bert_2018}. The BERT multilingual base model was trained on multilingual Wikipedia data (102 languages). As SSCI-SciBERT was trained only for English it was not applicable for the fine-tuning with multilingual data. Therefore, we utilized a multilingual language representation model for the fine-tuning.
The model was fine-tuned using a dataset of 14,178 abstracts from scientific articles retrieved from the BASE and POLLUX collections of scientific articles. Abstracts from scientific articles in 3 languages (English, German and French) were selected for the training. 

The model is integrated into the soft filter of the BASE filtering pipeline in the POLLUX infrastructure. Nevertheless, the model can be used separately with the Transformers library, as Listing~\ref{listing:bert-base-ml-politics} demonstrates.  

\begin{minipage}{\linewidth}
\begin{lstlisting}[caption=Usage example of bert-base-ml-politics with the Transformers library, label = listing:bert-base-ml-politics]
from transformers import AutoTokenizer, AutoModelForSequenceClassification
from transformers import pipeline

tokenizer = AutoTokenizer.from_pretrained('kalawinka/bert-base-ml-politics')
model = AutoModelForSequenceClassification.from_pretrained('kalawinka/bert-base-ml-politics')
pipe = pipeline("text-classification", model=model, tokenizer = tokenizer, max_length=512, truncation=True)

pipe('add scientific abstract')
\end{lstlisting}
\end{minipage}

Applying this model to the abstract of this technical report produces the output illustrated in Listing~\ref{bert-base-ml-politics-out}.

\begin{minipage}{\linewidth}
\begin{lstlisting}[caption = Example of bert-base-ml-politics output, label = bert-base-ml-politics-out]
    [{'label': 'politics', 'score': 0.9972042441368103}]
\end{lstlisting}
\end{minipage}

As SSciBERT\_politics, the model was finetuned using the Transformers library as Listing~\ref{listing:bert-base-ml-politics-training} demonstrates. The AdamW algorithm was applied to minimise the loss function. Evaluation was done at the end of each epoch. The total number of training epochs to perform was set to 3. The initial learning rate was set to 5e-5. The training was performed using one NVIDIA A40 GPU. Training time comprises 52,20334 minutes.

\begin{minipage}{\linewidth}
\begin{lstlisting}[caption = Training parameters for the finetuning  of bert-base-ml-politics with the Transformers library, label = listing:bert-base-ml-politics-training]
training_args = TrainingArguments(output_dir=output_dir, 
                                  evaluation_strategy="epoch",
                                  num_train_epochs = 3,
                                  )
\end{lstlisting}
\end{minipage}

\subsubsection{Training data}

\begin{figure}[h!]
\centering
  \includegraphics[width=1\textwidth]{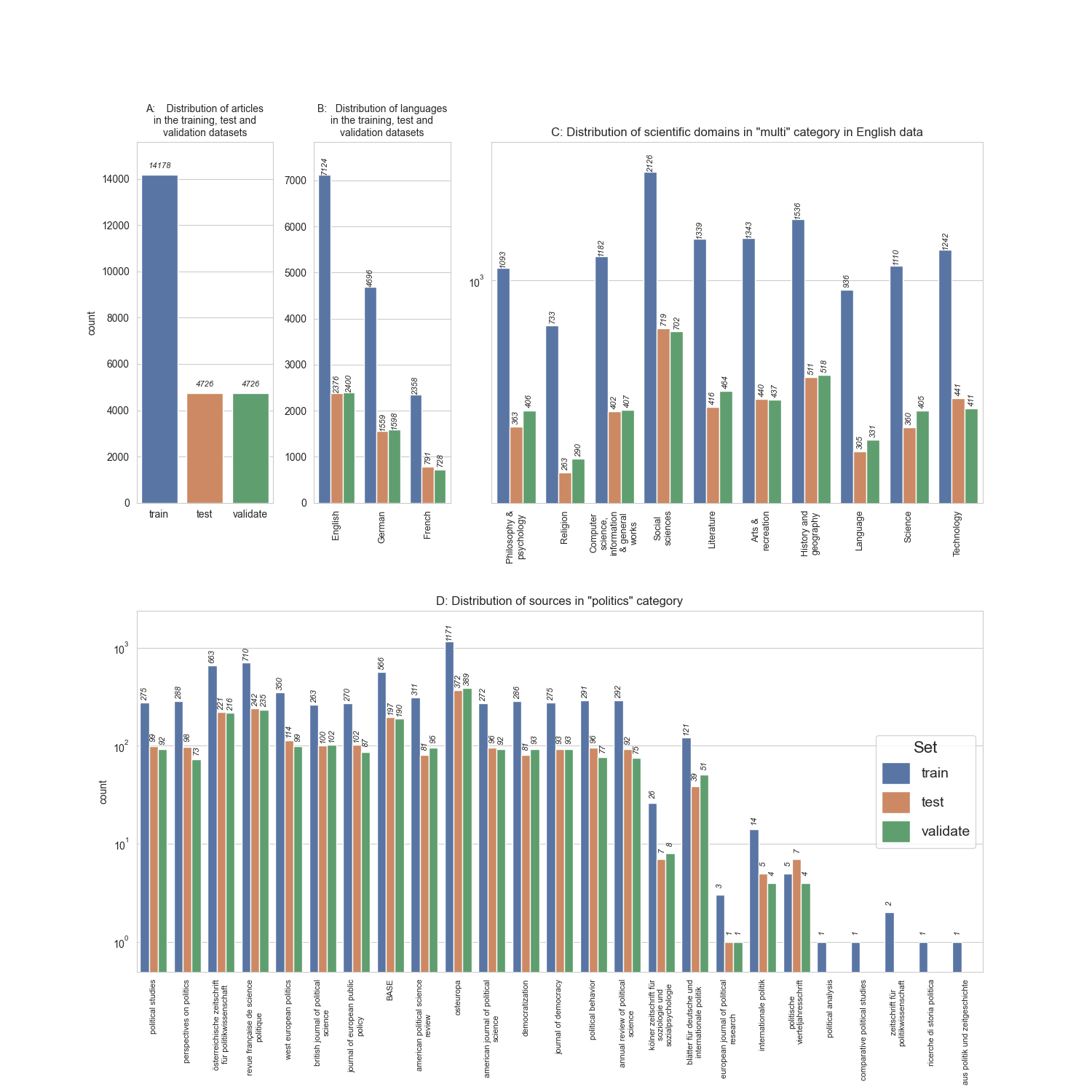}
  \caption{Distribution of training data in the multilingual corpus}
  \label{fig:corpus2_combined}
  %\vspace{-4mm}
\end{figure}

The collection of the training data was similar to the one described in Section~\ref{training_data_english}. Data from the BASE and POLLUX collections were utilized to construct the training corpus. A semi-automated method was employed to generate the labelled training dataset.
To obtain scientific abstracts from the political science domain we used both POLLUX and BASE collection. We created a list of political science journals from German-, English- and French-speaking countries. Afterwards, we collected articles from the POLLUX collection published in these journals containing an abstract and publication date. In the next step, we collected articles from the BASE collection corresponding to the following criteria. Only articles with the types article or review, DDC classification 320-328 (Political science), written in English, German or French, containing an abstract were selected. Figure~\ref{fig:corpus2_combined}-D shows the distribution of sources in the 'politics' category. To collect data from other (not political science) domains, we utilized only articles from the BASE collection. We sampled articles which correspond to the following criteria. Only articles with the types article or review, DDC classification not 320-328 (Political science), written in English, German or French, containing an abstract were selected Additionally we designed our query to restrict the selection only to articles which do not contain the keyword 'politi*' in title, keywords or abstracts. Figure~\ref{fig:corpus2_combined}~-C shows the distribution of scientific domains in the 'multi' category\footnote{One article can belong to multiple scientific domains.}. After collecting the data from both collections, we verified the language of the abstracts to ensure they were in English, German, or French and excluded abstracts containing fewer than 20 words.   

Figure~\ref{fig:corpus2_combined}-A shows the distribution of articles in the training, test and validation corpora. Thus, test and validation datasets contain 20\% each and the training corpus comprises 60\% of all data. Figure~\ref{fig:corpus2_combined}-B shows the distribution of languages in the training data. English is a prevailing language, as we encountered difficulties locating sufficient data in German and French. Furthermore, our training data is limited to these languages because we were unable to find adequate training data in other languages that met our criteria. Specifically, we required data that included both a title and a valid abstract. Many multilingual entries contained abstracts only in English. Additionally, for the BASE dataset, we required entries with the DDC classification. 

%Figures~\ref{fig:corpus2_combined_en}, \ref{fig:corpus2_combined_de}, and \ref{fig:corpus2_combined_fr} show the distribution of articles in the training, test and validation datasets (subplot A), distribution of scientific domains in the 'multi' category (subplot B) and the distribution of sources in the 'politics' category in English, German, and French subsets subsequently. 

%\subsubsection{Training}

\section{Evaluation}

% to do: rephrase. add info about sscibert

 We evaluated both approaches using an annotated dataset of 4,726 abstracts from different scientific domains. The weighted keyword-based approach was tested in two modes: applied to title, keywords and abstract, and applied only to title and keywords. The SSciBERT\_politics model was evaluated on a separate dataset consisting of 973 abstracts written in English.

Table~\ref{tab:eval_clas} shows an evaluation of the approaches by class for all languages. Keyword-based approach, applied to title, abstract and keywords showed a high accuracy of 0,87, as well as a high F1-score for both classes: multi and politics. The keyword-based approach applied only to titles and keywords showed a slightly deteriorating performance of 0,8. Class politics has high precision and low recall, at the same time class multi has high recall and low precision.
Both BERT-based classifiers showed high total accuracy and F1-scores for all classes. The bert-base-ml-politics showed higher accuracy (0.978) than the SSciBERT\_politics model (0.9).

\begin{table}[h!]
\begin{tabular}{llrrrrr}
\toprule
                                  approach &    label &  precision &  recall &  f1-score &  support &  accuracy \\
\midrule
keyword filter (title, abstract, keywords) & politics &      0.871 &   0.846 &     0.859 &     2143 &     \\
keyword filter (title, abstract, keywords) &    multi &      0.876 &   0.896 &     0.886 &     2583 &       \textbf{0.874} \\
\midrule
          keyword filter (title, keywords) & politics &      \textbf{\textit{0.947}} &   \textbf{\textit{0.593}} &     0.729 &     2143 &      \\
          keyword filter (title, keywords) &    multi &      \textbf{\textit{0.742}} &   \textbf{\textit{0.973}} &     0.842 &     2583 &      \textbf{0.800} \\
          \midrule
                     bert-base-ml-politics & politics &      0.975 &   0.978 &     0.976 &     2143 &     \\
                     bert-base-ml-politics &    multi &      0.981 &   0.979 &     0.980 &     2583 &       \textbf{ 0.978} \\
                     \midrule
                         SSciBERT\_politics & politics &      0.889 &   0.902 &     0.895 &      460 &      \\
                         SSciBERT\_politics &    multi &      0.911 &   0.899 &     0.905 &      513 &       \textbf{0.900 }\\
\bottomrule
\end{tabular}
\caption{Evaluation by class}
    \label{tab:eval_clas}
\end{table}

Table~\ref{tab:eval_class_lang} presents the evaluation results by class and language, which align with the findings in Table Table~\ref{tab:eval_clas}. The bert-base-ml-politics model demonstrates the best performance across all languages. Conversely, the keyword-based approach applied to titles and keywords exhibits low performance for French, with an F1-score of 0.343. However, French had the least amount of training and evaluation data within the entire corpus, potentially introducing bias into the evaluation for all algorithms

\begin{table}[h!]
\begin{tabular}{lllrrrr}
\toprule
                                  approach &    label & language &  precision &  recall &  f1-score &  support \\
\midrule
keyword filter (title, abstract, keywords) & politics &  English &      0.931 &   0.839 &     0.883 &     1212 \\
keyword filter (title, abstract, keywords) &    multi &  English &      0.848 &   0.936 &     0.890 &     1164 \\
keyword filter (title, abstract, keywords) & politics &   German &      0.843 &   0.861 &     0.852 &      783 \\
keyword filter (title, abstract, keywords) &    multi &   German &      0.856 &   0.838 &     0.847 &      776 \\
keyword filter (title, abstract, keywords) & politics &   French &      0.647 &   0.831 &     0.728 &      148 \\
keyword filter (title, abstract, keywords) &    multi &   French &      0.958 &   0.896 &     0.926 &      643 \\
\midrule
          keyword filter (title, keywords) & politics &  English &      0.976 &   0.568 &     0.718 &     1212 \\
          keyword filter (title, keywords) &    multi &  English &      0.686 &   0.985 &     0.809 &     1164 \\
          keyword filter (title, keywords) & politics &   German &      0.943 &   0.699 &     0.803 &      783 \\
          keyword filter (title, keywords) &    multi &   German &      0.759 &   0.957 &     0.847 &      776 \\
          keyword filter (title, keywords) & politics &   French &     \textbf{\textit{ 0.625}} &   \textbf{\textit{0.236}} &     \textbf{\textit{0.343}} &      148 \\
          keyword filter (title, keywords) &    multi &   French &      0.846 &   0.967 &     0.903 &      643 \\
          \midrule
                     bert-base-ml-politics & politics &  English &      0.989 &   0.993 &     0.991 &     1212 \\
                     bert-base-ml-politics &    multi &  English &      0.992 &   0.989 &     0.991 &     1164 \\
                     bert-base-ml-politics & politics &   German &      0.952 &   0.958 &     0.955 &      783 \\
                     bert-base-ml-politics &    multi &   German &      0.957 &   0.951 &     0.954 &      776 \\
                     bert-base-ml-politics & politics &   French &      0.979 &   0.959 &     0.969 &      148 \\
                     bert-base-ml-politics &    multi &   French &      0.991 &   0.995 &     0.993 &      643 \\
\bottomrule
\end{tabular}
    \caption{Evaluation by class and language}
    \label{tab:eval_class_lang}
\end{table}

\section{Conclusion}\label{sec:conc}

In general, the BERT-based classifier demonstrated the highest accuracy, but at the same time the highest processing time, approximately 4.5 seconds (on a CPU) to process a single abstract. Conversely, the keyword filter, although slightly less accurate, exhibited significantly faster performance compared to the classification model (less than 1 second to process a single abstract). The efficacy of the keyword filter improves when applied to all available metadata.

Both methodologies are applicable to filtering political science data. The keyword-based approach is particularly advantageous when dealing with large datasets and limited computational resources for running complex models. Furthermore, this approach is also suitable for processing articles with incomplete metadata.

The main limitation of the method is a lack of manually annotated training data. To address this issue we applied a semi-automated labelling approach. The proposed filtering approach can be applied for filtering metadata from other scientific domains and therefore improve the overview of the domain-related literature and facilitate efficiency in research. 

\section*{Acknowledgements}
Nina Smirnova acknowledges support by Deutsche Forschungsgemeinschaft (DFG) under grant number MA 3964/7-2, the Fachinformationsdienst Politikwissenschaft -- Pollux. The core BASE-filtering pipeline was developed by the team of the State and University Library Bremen (SuUB): Martin Blenkle, Elmar Haake, Thore Christiansen, Marie-Saphira Flug, Lena Klaproth. The weighted keyword-based filtering approach was developed by Daniel Opitz. Michael Czolkoß-Hettwer contributed to the creation of the list of keywords for the political science domain. Nina Smirnova was founded by the European Union under the Horizon Europe OMINO - Overcoming Multilevel Information Overload (101086321, https://ominoproject.eu/). Views and opinions expressed are those of the authors alone and do not necessarily reflect those of the European Union or the European Research Executive Agency.  Neither the European Union nor the European Research Executive Agency can be held responsible for them.

\clearpage
\bibliographystyle{unsrtnat}
\bibliography{references}  
\end{document}